\documentclass[a4paper]{article}
\usepackage{cite}
\usepackage{INTERSPEECH2020}

\usepackage{amssymb}
\newcommand{\argmax}{\mathop{\rm arg~max}\limits}

\usepackage{colortbl}
\newcolumntype{t}{!{\vrule width 0.1pt}} 
 
\newcolumntype{b}{!{\vrule width 1.5pt}}

\title{Exploring Optimal DNN Architecture for End-to-End Beamformers Based on Time-frequency References}
\name{Yuichiro Koyama$^{12}$, Bhiksha Raj$^1$}
\address{
  $^1$Carnegie Mellon University, Pittsburgh, PA, USA\\
  $^2$ Sony Corporation, Tokyo, Japan}
\email{Yuichiro.Koyama@sony.com, bhiksha@cs.cmu.edu}

\begin{document}
\setlength{\abovedisplayskip}{5pt} 
\setlength{\belowdisplayskip}{5pt} 

\maketitle

\begin{abstract}
Acoustic beamformers have been widely used to enhance audio signals. Currently, the best methods are the deep neural network (DNN)-powered variants of the generalized eigenvalue and minimum-variance distortionless response beamformers and the DNN-based filter-estimation methods that are used to directly compute beamforming filters. Both approaches are effective; however, they have blind spots in their generalizability. Therefore, we propose a novel approach for combining these two methods into a single framework that attempts to exploit the best features of both. The resulting model, called the W-Net beamformer, includes two components; the first computes time-frequency references that the second uses to estimate beamforming filters. The results on data that include a wide variety of room and noise conditions, including static and mobile noise sources, show that the proposed beamformer outperforms other methods on all tested evaluation metrics, which signifies that the proposed architecture allows for effective computation of the beamforming filters.
\end{abstract}


\section{Introduction}
\label{sec:intro}
The acoustic beamformer, which combines signals captured by an array of microphones to produce an enhanced signal, is widely used for various applications, such as 
automatic speech recognition (ASR), speaker recognition, and hearing aids~\cite{barker2015third,vincent2017analysis,barker2018fifth,movsner2018dereverberation,jensen2015analysis}. The quest for improved beamformer algorithms remains an active area of research. 

Traditional beamformer algorithms required accurate preliminary detection of the direction of target sources~\cite{griffiths1982alternative, capon1969high, bitzer2001superdirective, frost1972algorithm};
this is a task fraught with difficulties. More recent methods have employed the generalized eigenvalue (GEV) beamformer~\cite{van1988beamforming,warsitz2007blind}, which is a special case of the generalized singular value decomposition-based algorithm~\cite{doclo2002gsvd,serizel2014low} and does not require explicit source localization, but instead directly ``beamforms'' to the direction of the highest signal to noise ratio (SNR). However, the need for explicit localization has been replaced by the equally challenging task of computing accurate cross-power spectral density matrices for noise.

Similar to the trend in several fields of artificial intelligence, this problem has been found to be amenable to solutions based on deep learning. The most empirically successful methods have been based on the {\em mask-estimation} (ME) approach, which estimates a {\em time-frequency mask} that localizes the spectro-temporal regions of the signal that are dominated by noise, to compute its cross-power spectral density~\cite{heymann2016neural, zhou2018robust, higuchi2018frame, wang2018mask, liu2018neural}. However, this approach has some drawbacks. The identification of ``noise'' and ``speech'' time-frequency components is partially based on heuristic thresholds; the estimators are then trained based on these ad hoc thresholds.
The performance of mask estimators may be improved by jointly optimizing them with acoustic models for ASR. It has been shown that such joint optimization achieves better ASR performance~\cite{ochiai2017multichannel,heymann2017beamnet,xu2019joint}. However, the obtained masks depend on the specific task (ASR) and are not appropriate for general purposes. 
Additionally, the computation of the spectral density matrices combines information from the entire recording, effectively assuming that the noise and source are spatially stationary. Hence, tracking mobile sources remains a challenge~\cite{Bddeker2018ExploringPA}.

An alternate deep neural network (DNN)-based approach, the {\em filter-estimation} (FE) approach, is employed to directly estimate the filters in a {\em filter-and-sum} beamformer~\cite{johnson1993array} using a DNN. This approach attempts to handle the spatio-temporal non-stationarity of signals through recurrent architectures that update filter parameters with time~\cite{xiao2016study, meng2017deep, pfeifenberger2019deep}. This method is considerably effective when test conditions are similar to those seen in training data. However, it lacks the significant robustness and generalization of the mask-based method, where the DNN only performs the substantially simpler task of ME and thus generalizes better.
The filter-and-sum network (FaSNet)~\cite{luo2019fasnet}, which is an FE approach based on a two-stage architecture and time-domain processing, shows its generalizability for the location of the sources.
However, only one type of architecture is evaluated in the study, and the rationale of its architecture is not investigated.

In this paper, we propose a hybrid beamforming solution that combines the most useful aspects of both the aforementioned approaches. 
Specifically, inspired by the pipeline of ME and FaSNet, 
we propose a two-stage model based on the time-frequency domain, where the first stage computes the time-frequency references that are then used by the second stage to compute the actual beamforming filters. Our method combines the superior generalization of the mask-based approach with the more optimal processing of the FE approach.
The contributions of this work are threefold. 
First, we prove that the performance of FE approaches tends to be higher than that of the ME approaches.
Second, our proposed methods adopt a two-stage architecture and consistently surpass the ME approaches and the baseline of the FE approach.
Third, we show that
a concatenation operation is more suitable than an attention mechanism for the joint two-stage architecture.

\section{Background and related works}
\label{sec:format}
We consider an array with $M$ microphones. The signal captured by the $m^{\rm th}$ microphone in the array can be written in the time-frequency domain as
\begin{equation}
X_m(t,f) = H_m(t,f)S(t,f) + N_m(t,f),
\end{equation}
where $S(t,f)$ represents the complex time-frequency coefficients of the clean speech
at time frame index $t$ and frequency bin $f$
, $N_m(t,f)$ represents the noise at the $m^{\rm th}$microphone, $H_m(f)$ is the room impulse response at the $m^{\rm th}$ microphone, and $X_m(t,f)$ is the noisy signal actually captured by the $m^{\rm th}$ microphone. The task of the beamformer is to combine the signals $X_m(t,f)$ in such a manner that the combined signal is as close to $S(t,f)$ or $H_r(t,f)S(t,f)$ as possible ($r$ represents a desired channel).

The most successful approach to beamforming is the {\em filter-and-sum} approach, in which each microphone signal is processed by a separate filter, and the filtered signals are subsequently summed up to obtain the enhanced signal as follows:
\begin{equation}
\hat{S}(t,f) = \bm{w}^H(f) \bm{x}(t,f),
\label{eq:FandS}
\end{equation}
where $\bm{w}(f) = [W_1(f), \cdots, W_M(f)]^\top$, $W_m(f)$ represents the filter applied to the $m^{\rm th}$ array signal, and $\bm{x}(t,f) = [X_1(t,f), \cdots, X_M(t,f)]^\top$. The challenge is to estimate the filter $\bm{w}(f)$ that results in the $\hat{S}(t,f)$ with the highest SNR.

\subsection{Traditional solutions}
Traditional beamforming solutions estimate $\bm{w}(f)$ to optimize various objective measures computed from the known location of the source (the ``look'' direction) and various criteria related to the SNR of the enhanced signal, such as the SNR under spatially uncorrelated noise~\cite{johnson1993array}, ratio of the energy from the look direction to that from other directions~\cite{bitzer2001superdirective}, minimum noise variance~\cite{capon1969high}, and minimization of signal energy from non-look directions~\cite{griffiths1982alternative,frost1972algorithm}. These methods generally require knowledge of the source direction and are sensitive to errors in estimating it.

\subsection{Mask-based beamformers}
The GEV beamformer~\cite{van1988beamforming,warsitz2007blind}, which is a special case of the generalized singular value decomposition-based algorithm~\cite{doclo2002gsvd,serizel2014low}, bypasses the requirement of knowledge of the look direction by estimating the filter parameters to maximize the signal to noise ratio as follows:
\begin{equation}
\text{SNR}(f) = \frac{\bm{w}^H(f) \bm{\Phi}_S(f) \bm{w}(f) }{\bm{w}^H(f) \bm{\Phi}_N(f) \bm{w}(f)},
\end{equation}
where $\bm{\Phi}_S(f)$ and $\bm{\Phi}_N(f)$ are the cross-power spectral matrices for speech and noise, respectively. Maximization of the above objective results in a GEV solution, and the GEV beamforming filter is obtained as
\begin{equation}
\bm{w}_{\text{GEV}}(f) = \argmax_{\bm{w}(f)}\text{SNR}(f).
\label{eq:snrmax}
\end{equation}
Recently, the minimum-variance distortionless response (MVDR) beamforming filter has also been 
formulated without a {\em steering vector} directed to the location of the source~\cite{souden2009optimal},
\begin{equation}
\bm{w}_{\text{MVDR}}(f)=\frac{1}{\text{tr}(\bm{\Phi}_{N}(f)^{-1}\bm{\Phi}_{S}(f))}\bm{\Phi}_{N}(f)^{-1}\bm{\Phi}_{S}(f)\bm{u},
\end{equation}
where $\bm{u}=[0,\dots,1,\dots,0]^\top$ is the $M$ dimensional vector to select the desired channel.
These solutions require estimation of $\bm{\Phi}_S(f)$ and $\bm{\Phi}_N(f)$, which requires prior knowledge of the speech- and noise-dominated time-frequency regions of the array signals.

\begin{figure}[htb]
  \centering
  \centerline{\includegraphics[width=7.0cm]{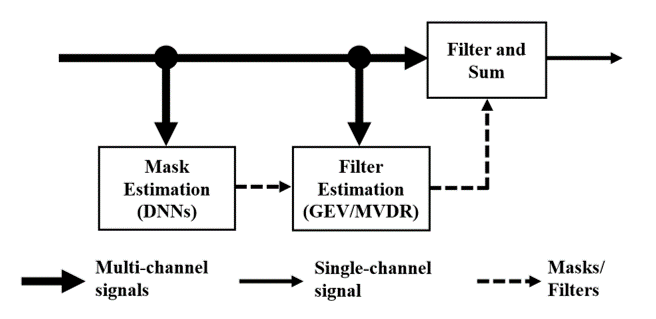}}
  \caption{Basic pipeline of mask-based beamformers}
\label{fig:basic_structure}
\end{figure}

The mask-based method~\cite{heymann2016neural, zhou2018robust, higuchi2018frame, wang2018mask, liu2018neural} solves this problem by estimating masks $r_N(t,f)$ and $r_S(t,f)$, which are computed using a DNN, and assigns to any time-frequency location $(t,f)$ the probability of being noise and speech dominated, respectively. Subsequently, the cross-power spectral matrices are computed as
\begin{equation}
\bm{\Phi}_S(f) = \frac{1}{T}\sum_{t=1}^{T}[r_S(t,f)\bm{x}(t,f)\bm{x}^H(t,f)],
\label{eq:mask_s}
\end{equation}
\begin{equation}
\bm{\Phi}_N(f) = \frac{1}{T}\sum_{t=1}^{T}[r_N(t,f)\bm{x}(t,f)\bm{x}^H(t,f)].
\label{eq:mask_n}
\end{equation}
The DNNs that compute $r_S(t,f)$ and $r_N(t,f)$ are trained from signals with known noise. A heuristically chosen threshold $\theta$ is applied to each time-frequency component of the training signals to decide whether to label it as noise or signal dominated, to train the DNN. Fig.\ref{fig:basic_structure} depicts the pipeline of mask-based beamformers, as mentioned previously.

\subsection{Beamformers with directly estimated filters}
The FE approach directly applies a DNN
to the multi-channel inputs to compute the time-varying beamforming filters $\bm{w}(t,f)$~\cite{xiao2016study, meng2017deep, pfeifenberger2019deep}. The network is trained on multi-channel signals, with targets provided either by a clean signal or by higher-level models such as an ASR system~\cite{meng2017deep}.

We have previously explained the relative merits and demerits of beamformer methods. Traditional methods require knowledge of the look direction and GEV and MVDR beamformers estimate a single $\bm{w}(f)$ for the entire signal. Moreover, while filter-based methods can estimate time-varying filters, they are biased towards noise and environments observed in training as no additional cues relating to the noise in a current test condition, except the noisy signal itself, are provided to the algorithms.
The method we propose addresses these drawbacks.

\section{Proposed method}
\label{sec:method}
\begin{figure*}[htb]
  \centering
  \includegraphics[width=16.0cm]{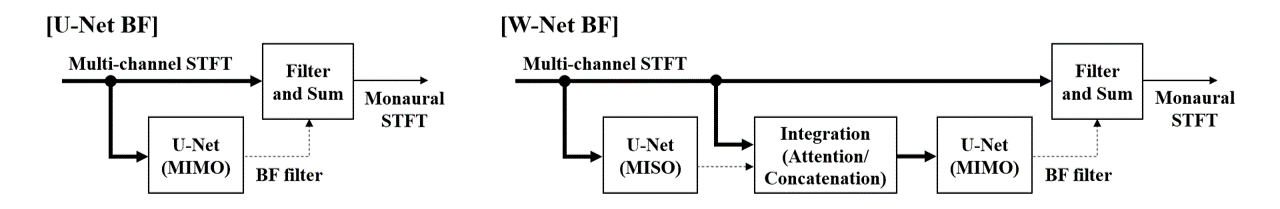}
  \caption{Structure of our proposed beamformers. We propose and compare three types of FE approaches: U-Net BF, W-Net BF with attention, and W-Net BF with concatenation.}
\label{fig:proposed_methods}
\vspace{-5mm}
\end{figure*}

Figure \ref{fig:proposed_methods} presents the structure of our proposed beamformers. 
We propose and compare three types of FE approaches utilizing a {\em U-Net}.
The choice of U-Net is predicated in part on the success of these networks in other image and speech processing tasks~\cite{ronneberger2015u,jansson2017singing}. 
The U-Net comprises a series of convolutional layers of decreasing size, which attempt to spatially ``summarize'' the input into a bottleneck, followed by a series of deconvolutional layers that reconstruct the target output from the bottleneck. The U-Net is ideally suited to problems such as denoising, where an output of the same spatial extent as the input must be derived while ``squeezing-out'' the noise in the input. For example, in~\cite{jansson2017singing}, it is demonstrated that the U-Net is suitable for separating speech and music signals.

All approaches are formulated as a problem of estimating $\mathbf{W}\in\mathbb{R}^{T\times F\times 2M}$ from $\mathbf{X}\in\mathbb{R}^{T\times F\times 2M}$, where $\mathbf{W}$ is a tensor representation of the time-variant beamforming filter $W^*_m(t,f)$ and $\mathbf{X}$ is a tensor representation of $X_m(t,f)$ in the range of $t=1,\dots,T$ and $f=1,\dots,F$.
One half of $\mathbf{X}$ ($1,\dots,M$) represents amplitude and the other represents the phase of the original complex value $X_m(t,f)$ 
Similarly, one half of $\mathbf{W}$ ($1,\dots,M$) represents the real portion and the other half represents the imaginary portion of the original complex value $W^*_m(t,f)$.
The enhanced signal $\hat{S}(t,f)$ is obtained using $\bm{w}(t,f) = [W_1(t,f), \cdots, W_M(t,f)]^\top$ as
\begin{equation}
\hat{S}(t,f) = \bm{w}^H(t,f) \bm{x}(t,f).
\label{eq:FandS_tv}
\end{equation}

\subsection{U-Net beamformer}
The first method we propose attempts to directly estimate beamforming filters using the multi-input-multi-output (MIMO) version of U-Net (MIMO U-Net) as
\begin{equation}
\mathbf{W} = \text{UNET}_{2M,2M}(\mathbf{X}),
\end{equation}
where $\text{UNET}_{p,q}$ represents the U-Net that 
can take $p$ channel time-frequency features as input and output $q$ channel time-frequency features.
We refer to this approach as a U-Net beamformer (U-Net BF), which is considered to be the baseline method of the FE approach.

\subsection{W-Net beamformer with attention}
Our second proposed method is based on a two-stage architecture that utilizes two types of U-Net: 
multi-input-single-output and MIMO.
The first estimates a ``reference'' time-frequency representation $\mathbf{Y}\in\mathbb{R}^{T\times F\times 1}$, which is expected to be helpful for computing the beamforming filter
as
\begin{equation}
\mathbf{Y} = \text{UNET}_{2M,1}(\mathbf{X}).
\end{equation}
The integration block combines this reference with the multi-channel input signals using an attention mechanism as
\begin{equation}
\mathbf{Z}_\text{attn} = \text{sigmoid}(\mathbf{Y}) \odot \mathbf{X},
\label{eq:attn}
\end{equation}
where $\mathbf{Z}_\text{attn}\in\mathbb{R}^{T\times F\times 2M}$
represents the integrated time-frequency features, the $\text{sigmoid}(\cdot)$ is an element-wise sigmoid function, and 
$\odot$ is the Hadamard product.
This integration can be interpreted as an analogy of time-frequency masking such as $r_S(t,f)$ and $r_N(t,f)$ in Eq.~\eqref{eq:mask_s}-\eqref{eq:mask_n}.
The obtained $\mathbf{Z}$ is then fed into the second U-Net, which estimates the time-varying beamforming filters as
\begin{equation}
\mathbf{W} = \text{UNET}_{2M,2M}(\mathbf{X}).
\end{equation}
We refer to this two-stage method as a W-Net beamformer (W-Net BF) with attention.

\subsection{W-Net beamformer with concatenation}
The only difference between the second and third methods is the integration block.
Instead of Eq.~\eqref{eq:attn}, a concatenation operation along the channel dimension is assigned to the integration block for the third method as
\begin{equation}
\mathbf{Z}_\text{concat} = \text{concatenation}(\mathbf{Y}, \mathbf{X}),
\end{equation}
where $\mathbf{Z}_\text{concat}\in\mathbb{R}^{T\times F\times 2M+1}$
is the concatenated time-frequency features fed into the second U-Net as
\begin{equation}
\mathbf{W} = \text{UNET}_{2M+1,2M}(\mathbf{X}).
\end{equation}
We refer to this two-stage method as a W-Net BF with concatenation.

\subsection{Training the U-Net and W-Net beamformers}
To train the network, we require collections of training array recordings and an additional ``noise-free reference'' channel (not to be confused with $\mathbf{Y}$). This reference channel corresponds to a recording that has been captured by a reference microphone that is influenced by the room impulse response, but not by noise. This is a common requirement for DNN-based beamformer algorithms. This may be recorded by a highly directional microphone used only during training, or a channel to which noise has not been added can be used for synthetic data. For this reference channel, which is  denoted by the subscript $R$, we have $S_{R}(t,f) = H_r(f)S(t,f)$, where $r=1$. 
We minimize the following loss function:
\begin{equation}
L= \frac{1}{TF}\sum_{t=1}^{T}\sum_{f=1}^F \|\hat{S}(t,f) - S_R(t,f)\|^2.
\label{eq:loss}
\end{equation}

\section{Experiments}
\label{sec:typestyle}
\begin{table*}
\begin{center}
    \caption{Evaluation result. The values of SI-SNR are represented in dB. The performance of the FE approaches tends to be higher than that of the ME approaches, and the W-Net beamformers surpass not only the ME approaches but also the ideal binary mask (IBM) in terms of all metrics.
The difference between the performances of the ME and FE approaches is larger in moving conditions.}

  \label{tab:results}
\scalebox{0.75}{
    \centering
    \begin{tabular}{lbccc|cccbccc|ccc|ccc}
\toprule
      & \multicolumn{6}{cb}{Anechoic} & \multicolumn{9}{c}{Reverberation} \\ 

      & \multicolumn{3}{c|}{static} & \multicolumn{3}{cb}{moving} & \multicolumn{3}{c|}{static} & \multicolumn{3}{c|}{moving} & 
      \multicolumn{3}{c}{ACE corpus} \\ 
\midrule

        Method & SI-SNR & STOI & PESQ & SI-SNR & STOI & PESQ & SI-SNR & STOI & PESQ & SI-SNR & STOI & PESQ & SI-SNR & STOI & PESQ \\ 
        \midrule
        Noisy & 4.88  & 0.81  & 1.45  & 4.87  & 0.81  & 1.46  & 4.94  & 0.77  & 1.54  & 4.87  & 0.77  & 1.53  & 5.95  & 0.92  & 2.08  \\
        \midrule
        BLSTM-GEV & 7.30  & 0.87  & 2.43  & 3.18  & 0.82  & 1.72  & 2.50  & 0.75  & 1.76  & -0.06  & 0.67  & 1.44  & -0.07  & 0.81  & 1.70  \\ 
        BLSTM-MVDR & 9.40  & 0.92  & 2.43  & 6.81  & 0.89  & 2.06  & 7.00  & 0.84  & 1.94  & 5.36  & 0.79  & 1.66  & 6.46  & 0.94  & 2.17  \\ 
        \midrule
        U-Net BF & 13.58  & 0.89  & 2.30  & 13.49  & 0.89  & 2.25  & 12.74  & 0.85  & 2.33  & 12.56  & 0.85  & 2.27  & 16.38  & \bf{0.95}  & 2.76  \\ 
        W-Net BF attn. & 19.55  & \bf{0.94}  & 2.91  & 18.57  & \bf{0.94}  & 2.78  & 13.06  & \bf{0.86}  & 2.44  & 12.83  & \bf{0.86}  & 2.37  & 16.44  & \bf{0.95}  & 2.82  \\ 
        W-Net BF concat. & \bf{20.04}  & \bf{0.94}  & \bf{2.98}  & \bf{18.98}  & \bf{0.94}  & \bf{2.84}  & \bf{13.41}  & \bf{0.86}  & \bf{2.50}  & \bf{13.18}  & \bf{0.86}  & \bf{2.42}  & \bf{16.56}  & \bf{0.95}  & \bf{2.85}  \\ 
        \midrule
        IBM & 17.75  & 0.93  & 2.80  & 17.77  & 0.93  & 2.80  & 16.91  & 0.93  & 3.01  & 16.85  & 0.93  & 3.00  & 20.03  & 0.98  & 3.51  \\ 
        \bottomrule
    \end{tabular}
    }
    \end{center}
\vspace{-6mm}
\end{table*}

\subsection{Dataset}
To evaluate the proposed method in sufficiently variable conditions, we constructed two sets of simulated microphone array recordings based on room impulse response simulations:
Anechoic (i.e., without reverberation) and 
Reverberation datasets, which are further divided into two conditions:
static and moving sources. 

For these datasets, we considered a rectangular room of dimensions $(l_x,l_y,l_z)$, with the origin at one corner.
A linear array with an aperture of 30.0 cm with six microphones ($M=6$) was located at a randomly determined point in the room.
The impulse responses from a randomly determined point in the room to the microphone array was generated using the image-source method~\cite{lehmann2008prediction}.
We randomly set $l_x,l_y,l_z$ in the range $l_x \in (3.0~\text{m}, 10.0~\text{m}), l_y \in (3.0~\text{m},8.0~\text{m}), l_z \in (2.5~\text{m}, 6.0~\text{m})$, and the reverberation time T60 
in the range $\text{T60} \in (0.2~\text{s},0.8~\text{s})$.
These impulse responses were split as 80\% for training, 10\% for validation, and 10\% for testing.
We used the Wall Street Journal (WSJ0) corpus~\cite{garofalo2007csr} for the speech source, \textit{si\_tr\_s} for training, \textit{si\_dt\_05} for validation, and \textit{si\_et\_05} for testing.
We also used ESC-50~\cite{piczak2015dataset} and regular white and pink noise for the noise source. These sources were split as 80\% for training, 10\% for validation, and 10\% for testing per sound class.
One speech signal was generated by convolving an impulse response with a speech source, and one to three noise signals were generated by simultaneously convolving impulse responses with noise sources.
In the condition for moving sources, each source was moved with a velocity $v$ randomly determined as $v \in (0.1~\text{m/s}, 3.0~\text{m/s})$.

\subsection{Training and testing}
We trained the network on the Anechoic and Reverberation datasets respectively, while the static and moving sources were mixed at the same rate.  

For training and validation, we randomly sampled the impulse response, speech source, and noise source independently. 
We simulated the data obtained at the microphone array using the {\em on-the-fly} method~\cite{erdogan2018investigations}, and 50,000 iterations are defined as 1 epoch.
The SNR was set to follow a normal distribution with $N(\text{5dB},(\text{5dB})^2)$. 
All our proposed networks were trained with the Adam optimizer~\cite{kingma2014adam} for 600 epochs, and the learning rate was set as 0.002.
The parameters that minimize validation error were chosen for evaluation.
We also trained the bidirectional long short-term memory (BLSTM)-based ME network in the same manner as~\cite{heymann2016neural}, and utilized it for GEV and MVDR beamformers.
The GEV beamformer requires
a corrective postfilter after beamforming to achieve a distortionless
response; both the GEV filter and the recommended postfilter were
implemented.

For testing, we simulated the data obtained at the microphone array using impulse responses, speech sources, and noise sources. A total of 3000 recordings were generated with each SNR of 0, 5, and 10~dB, and the same data were utilized for all methods.
Furthermore, 
we utilized the Acoustic Characterization of Environments (ACE) corpus~\cite{eaton2016estimation} to evaluate how the methods trained in the Reverberation dataset could be generalized to actual measured data.
The ACE corpus is composed of impulse responses and noise measured by various microphone arrays.
We adopted an 8-channel linear array and extracted channels 2--7.
The speech signals in \textit{si\_et\_05} of the WSJ0 corpus were convolved with the measured impulse responses and mixed with the {\em ambient} and {\em fan} noises in the ACE corpus.

All recordings were sampled to 16 kHz. We computed a 1024-point short-time Fourier transform with a 75\% overlap and ignored the DC component, such that $F = 512$. Although the proposed architecture is fully convolutional and $T$ can be set to an arbitrary length, $T$ was set to 256 in the training phase.
For the U-Net block, the six convolution layers used kernels of size $3 \times 3$ with a stride of one. Between layers, rectified linear unit (ReLU) activations and an average pooling operation of size $2\times 2$ were performed over the blocks to decrease the spatial span of the input by half in each dimension. 
The six deconvolution layers used kernels of size $2 \times 2$ with a stride of 2 and the ReLU activations were performed. 
Batch normalization was performed after each convolution and deconvolution layer.
In the W-Net BF, the numbers of output channels of convolution and deconvolution were 16, 32, 64, 128, 256, 512, 256, 128, 64, 32, 16, and 1 in the first U-Net block.  
In the second U-Net block, the corresponding values were 16, 32, 64, 128, 256, 512, 256, 128, 64, 32, 16, and 12 ($2M$), resulting in 4.9 million parameters. 
In the U-Net BF, the numbers of output channels of convolution and deconvolution were set as 22, 45, 90, 180, 360, 720, 360, 180, 90, 45, 22, and 12 ($2M$), resulting in 4.84 million parameters. Thus,
the total number of parameters in the U-Net BF and W-Net BF were
comparable.

For evaluation, the test data were processed by the models and the performance was quantified using three metrics: average value of scale-invariant signal-to-noise ratio (SI-SNR)~\cite{roux2019sdr}, short-time objective intelligibility measure (STOI)~\cite{taal2011algorithm}, and perceptual evaluation of speech quality (PESQ)~\cite{rix2001perceptual}. 
To calculate these metrics, 
$S_R(t,f)$ was utilized as a reference signal.

\subsection{Results}
\label{subsec:results}
The evaluation results are shown in Table \ref{tab:results}. 
All the numbers are averages in terms of 3000 utterances and SNR conditions (i.e., 0, 5, 10 dB).
For comparison, we evaluate the ideal binary mask (IBM), which is an oracle approach for single-channel speech enhancement. 

We first discuss the result of the Anechoic dataset.
The performance of the FE approaches tends to be higher than that of the ME approaches, and the W-Net beamformers surpass both the ME approaches and the IBM for all metrics.
In addition, the difference between the performance of the ME approaches and that of the FE approaches is more significant in moving conditions, thereby showing that the time-varying beamforming filters from the FE approaches are better at handling spatially non-stationary noise sources.

Next, we discuss the results of the Reverberation dataset.
A tendency similar to the Anechoic dataset can be found in the static and moving conditions.
The performance of the FE approaches on the ACE corpus is considerably high. This can be interpreted as indicating that our proposed methods can obtain sufficient generalizability through training without actual measured impulse responses.

The results from both datasets show that the W-Net BF consistently outperforms the U-Net BF, which proves that the architecture of the W-Net BF allows for effective computation of the beamforming filters. 
Moreover, the W-Net BF with concatenation consistently outperforms the W-Net BF with attention, which shows that the concatenation operation along the channel dimension is an effective method to integrate the information based on time-frequency representation.

\section{Conclusion}
\label{sec:conclusion}
We proposed a novel DNN-based beamformer approach called the W-Net BF, which combines the best features of the ME and FE beamformers. 
Our proposed method combines a reference estimation module inspired by the former, and a beamforming filter estimation module inspired by the latter. 
Comparative evaluations showed that W-Net BF outperforms both approaches, which means that its architecture allows for effective computation of the beamforming filters.
It was also found that our proposed methods can obtain sufficient generalizability to deal with real-world observed signals.
In future work, we expect to expand this approach to other speech signal processing tasks. 
We will also explore joint training with specific models for applications such as ASR and 
speaker recognition systems.

\bibliographystyle{IEEEtran}

\bibliography{mybib}

\begin{thebibliography}{10}
\providecommand{\url}[1]{#1}
\csname url@samestyle\endcsname
\providecommand{\newblock}{\relax}
\providecommand{\bibinfo}[2]{#2}
\providecommand{\BIBentrySTDinterwordspacing}{\spaceskip=0pt\relax}
\providecommand{\BIBentryALTinterwordstretchfactor}{4}
\providecommand{\BIBentryALTinterwordspacing}{\spaceskip=\fontdimen2\font plus
\BIBentryALTinterwordstretchfactor\fontdimen3\font minus
  \fontdimen4\font\relax}
\providecommand{\BIBforeignlanguage}[2]{{%
\expandafter\ifx\csname l@#1\endcsname\relax
\typeout{** WARNING: IEEEtran.bst: No hyphenation pattern has been}%
\typeout{** loaded for the language `#1'. Using the pattern for}%
\typeout{** the default language instead.}%
\else
\language=\csname l@#1\endcsname
\fi
#2}}
\providecommand{\BIBdecl}{\relax}
\BIBdecl

\bibitem{barker2015third}
J.~{Barker}, R.~{Marxer}, E.~{Vincent}, and S.~{Watanabe}, ``The third
  ‘chime’ speech separation and recognition challenge: Dataset, task and
  baselines,'' in \emph{2015 IEEE Workshop on Automatic Speech Recognition and
  Understanding (ASRU)}, 2015, pp. 504--511.

\bibitem{vincent2017analysis}
E.~Vincent, S.~Watanabe, A.~A. Nugraha, J.~Barker, and R.~Marxer, ``An analysis
  of environment, microphone and data simulation mismatches in robust speech
  recognition,'' \emph{Computer Speech \& Language}, vol.~46, pp. 535--557,
  2017.

\bibitem{barker2018fifth}
J.~Barker, S.~Watanabe, E.~Vincent, and J.~Trmal, ``The fifth'chime'speech
  separation and recognition challenge: Dataset, task and baselines,''
  \emph{arXiv preprint arXiv:1803.10609}, 2018.

\bibitem{movsner2018dereverberation}
L.~{Mo^^c5^^a1ner}, P.~{Mat^^c4^^9bjka}, O.~{Novotn^^c3^^bd}, and J.~H.
  {^^c4^^8cernock^^c3^^bd}, ``Dereverberation and beamforming in far-field
  speaker recognition,'' in \emph{2018 IEEE International Conference on
  Acoustics, Speech and Signal Processing (ICASSP)}, 2018, pp. 5254--5258.

\bibitem{jensen2015analysis}
J.~{Jensen} and M.~S. {Pedersen}, ``Analysis of beamformer directed
  single-channel noise reduction system for hearing aid applications,'' in
  \emph{2015 IEEE ICASSP}, 2015, pp. 5728--5732.

\bibitem{griffiths1982alternative}
L.~{Griffiths} and C.~{Jim}, ``An alternative approach to linearly constrained
  adaptive beamforming,'' \emph{IEEE Transactions on Antennas and Propagation},
  vol.~30, no.~1, pp. 27--34, 1982.

\bibitem{capon1969high}
J.~{Capon}, ``High-resolution frequency-wavenumber spectrum analysis,''
  \emph{Proceedings of the IEEE}, vol.~57, no.~8, pp. 1408--1418, 1969.

\bibitem{bitzer2001superdirective}
J.~Bitzer and K.~U. Simmer, ``Superdirective microphone arrays,'' in
  \emph{Microphone arrays}.\hskip 1em plus 0.5em minus 0.4em\relax Springer,
  2001, pp. 19--38.

\bibitem{frost1972algorithm}
O.~L. {Frost}, ``An algorithm for linearly constrained adaptive array
  processing,'' \emph{Proceedings of the IEEE}, vol.~60, no.~8, pp. 926--935,
  1972.

\bibitem{van1988beamforming}
B.~D. {Van Veen} and K.~M. {Buckley}, ``Beamforming: a versatile approach to
  spatial filtering,'' \emph{IEEE ASSP Magazine}, vol.~5, no.~2, pp. 4--24,
  1988.

\bibitem{warsitz2007blind}
E.~{Warsitz} and R.~{Haeb-Umbach}, ``Blind acoustic beamforming based on
  generalized eigenvalue decomposition,'' \emph{IEEE Transactions on Audio,
  Speech, and Language Processing (TASLP)}, vol.~15, no.~5, pp. 1529--1539,
  2007.

\bibitem{doclo2002gsvd}
S.~{Doclo} and M.~{Moonen}, ``Gsvd-based optimal filtering for single and
  multimicrophone speech enhancement,'' \emph{IEEE Transactions on Signal
  Processing}, vol.~50, no.~9, pp. 2230--2244, 2002.

\bibitem{serizel2014low}
R.~{Serizel}, M.~{Moonen}, B.~{Van Dijk}, and J.~{Wouters}, ``Low-rank
  approximation based multichannel wiener filter algorithms for noise reduction
  with application in cochlear implants,'' \emph{IEEE/ACM TASLP}, vol.~22,
  no.~4, pp. 785--799, 2014.

\bibitem{heymann2016neural}
J.~{Heymann}, L.~{Drude}, and R.~{Haeb-Umbach}, ``Neural network based spectral
  mask estimation for acoustic beamforming,'' in \emph{2016 IEEE ICASSP}, 2016,
  pp. 196--200.

\bibitem{zhou2018robust}
Y.~{Zhou} and Y.~{Qian}, ``Robust mask estimation by integrating neural
  network-based and clustering-based approaches for adaptive acoustic
  beamforming,'' in \emph{2018 IEEE ICASSP}, 2018, pp. 536--540.

\bibitem{higuchi2018frame}
T.~{Higuchi}, K.~{Kinoshita}, N.~{Ito}, S.~{Karita}, and T.~{Nakatani},
  ``Frame-by-frame closed-form update for mask-based adaptive mvdr
  beamforming,'' in \emph{2018 IEEE ICASSP}, 2018, pp. 531--535.

\bibitem{wang2018mask}
Z.~{Wang} and D.~{Wang}, ``Mask weighted stft ratios for relative transfer
  function estimation and its application to robust asr,'' in \emph{2018 IEEE
  ICASSP}, 2018, pp. 5619--5623.

\bibitem{liu2018neural}
Y.~{Liu}, A.~{Ganguly}, K.~{Kamath}, and T.~{Kristjansson}, ``Neural network
  based time-frequency masking and steering vector estimation for two-channel
  mvdr beamforming,'' in \emph{2018 IEEE ICASSP}, 2018, pp. 6717--6721.

\bibitem{ochiai2017multichannel}
T.~Ochiai, S.~Watanabe, T.~Hori, and J.~R. Hershey, ``Multichannel end-to-end
  speech recognition,'' in \emph{Proceedings of the 34th International
  Conference on Machine Learning-Volume 70}.\hskip 1em plus 0.5em minus
  0.4em\relax JMLR. org, 2017, pp. 2632--2641.

\bibitem{heymann2017beamnet}
J.~{Heymann}, L.~{Drude}, C.~{Boeddeker}, P.~{Hanebrink}, and R.~{Haeb-Umbach},
  ``Beamnet: End-to-end training of a beamformer-supported multi-channel asr
  system,'' in \emph{2017 IEEE ICASSP}, 2017, pp. 5325--5329.

\bibitem{xu2019joint}
Y.~{Xu}, C.~{Weng}, L.~{Hui}, J.~{Liu}, M.~{Yu}, D.~{Su}, and D.~{Yu}, ``Joint
  training of complex ratio mask based beamformer and acoustic model for noise
  robust asr,'' in \emph{2019 IEEE ICASSP}, 2019, pp. 6745--6749.

\bibitem{Bddeker2018ExploringPA}
C.~{Boeddeker}, H.~{Erdogan}, T.~{Yoshioka}, and R.~{Haeb-Umbach}, ``Exploring
  practical aspects of neural mask-based beamforming for far-field speech
  recognition,'' in \emph{2018 IEEE ICASSP}, 2018, pp. 6697--6701.

\bibitem{johnson1993array}
D.~H. Johnson and D.~E. Dudgeon, \emph{Array signal processing: concepts and
  techniques}.\hskip 1em plus 0.5em minus 0.4em\relax PTR Prentice Hall
  Englewood Cliffs, 1993.

\bibitem{xiao2016study}
X.~Xiao, C.~Xu, Z.~Zhang, S.~Zhao, S.~Sun, S.~Watanabe, L.~Wang, L.~Xie, D.~L.
  Jones, E.~S. Chng \emph{et~al.}, ``A study of learning based beamforming
  methods for speech recognition,'' in \emph{CHiME 2016 workshop}, 2016, pp.
  26--31.

\bibitem{meng2017deep}
Z.~{Meng}, S.~{Watanabe}, J.~R. {Hershey}, and H.~{Erdogan}, ``Deep long
  short-term memory adaptive beamforming networks for multichannel robust
  speech recognition,'' in \emph{2017 IEEE ICASSP}, 2017, pp. 271--275.

\bibitem{pfeifenberger2019deep}
L.~{Pfeifenberger}, M.~{Z^^c3^^b6hrer}, and F.~{Pernkopf}, ``Deep
  complex-valued neural beamformers,'' in \emph{2019 IEEE ICASSP}, 2019, pp.
  2902--2906.

\bibitem{luo2019fasnet}
Y.~{Luo}, C.~{Han}, N.~{Mesgarani}, E.~{Ceolini}, and S.~{Liu}, ``Fasnet:
  Low-latency adaptive beamforming for multi-microphone audio processing,'' in
  \emph{2019 IEEE ASRU}, 2019, pp. 260--267.

\bibitem{souden2009optimal}
M.~{Souden}, J.~{Benesty}, and S.~{Affes}, ``On optimal frequency-domain
  multichannel linear filtering for noise reduction,'' \emph{IEEE TASLP},
  vol.~18, no.~2, pp. 260--276, 2010.

\bibitem{ronneberger2015u}
O.~Ronneberger, P.~Fischer, and T.~Brox, ``U-net: Convolutional networks for
  biomedical image segmentation,'' in \emph{International Conference on Medical
  image computing and computer-assisted intervention}.\hskip 1em plus 0.5em
  minus 0.4em\relax Springer, 2015, pp. 234--241.

\bibitem{jansson2017singing}
A.~Jansson, E.~Humphrey, N.~Montecchio, R.~Bittner, A.~Kumar, and T.~Weyde,
  ``Singing voice separation with deep u-net convolutional networks,'' in
  \emph{In Proceedings of the International Society for Music Information
  Retrieval Conference}, 2017, pp. 323--332.

\bibitem{lehmann2008prediction}
E.~A. Lehmann and A.~M. Johansson, ``Prediction of energy decay in room impulse
  responses simulated with an image-source model,'' \emph{The Journal of the
  Acoustical Society of America}, vol. 124, no.~1, pp. 269--277, 2008.

\bibitem{garofalo2007csr}
J.~Garofalo, D.~Graff, D.~Paul, and D.~Pallett, ``Csr-i (wsj0) complete,''
  \emph{Linguistic Data Consortium, Philadelphia}, 2007.

\bibitem{piczak2015dataset}
\BIBentryALTinterwordspacing
K.~J. Piczak, ``{ESC}: {Dataset} for {Environmental Sound Classification},'' in
  \emph{Proceedings of the 23rd {Annual ACM Conference} on {Multimedia}}.\hskip
  1em plus 0.5em minus 0.4em\relax {ACM Press}, 2015, pp. 1015--1018. [Online].
  Available: \url{http://dl.acm.org/citation.cfm?doid=2733373.2806390}
\BIBentrySTDinterwordspacing

\bibitem{erdogan2018investigations}
H.~Erdogan and T.~Yoshioka, ``Investigations on data augmentation and loss
  functions for deep learning based speech-background separation.'' in
  \emph{Interspeech}, 2018, pp. 3499--3503.

\bibitem{kingma2014adam}
D.~P. Kingma and J.~Ba, ``Adam: A method for stochastic optimization,''
  \emph{arXiv preprint arXiv:1412.6980}, 2014.

\bibitem{eaton2016estimation}
J.~{Eaton}, N.~D. {Gaubitch}, A.~H. {Moore}, and P.~A. {Naylor}, ``Estimation
  of room acoustic parameters: The ace challenge,'' \emph{IEEE/ACM TASLP},
  vol.~24, no.~10, pp. 1681--1693, 2016.

\bibitem{roux2019sdr}
J.~L. {Roux}, S.~{Wisdom}, H.~{Erdogan}, and J.~R. {Hershey}, ``Sdr
  ^^e2^^80^^93 half-baked or well done?'' in \emph{2019 IEEE ICASSP}, 2019, pp.
  626--630.

\bibitem{taal2011algorithm}
C.~H. {Taal}, R.~C. {Hendriks}, R.~{Heusdens}, and J.~{Jensen}, ``An algorithm
  for intelligibility prediction of time^^e2^^80^^93frequency weighted noisy
  speech,'' \emph{IEEE TASLP}, vol.~19, no.~7, pp. 2125--2136, 2011.

\bibitem{rix2001perceptual}
A.~W. {Rix}, J.~G. {Beerends}, M.~P. {Hollier}, and A.~P. {Hekstra},
  ``Perceptual evaluation of speech quality (pesq)-a new method for speech
  quality assessment of telephone networks and codecs,'' in \emph{2001 IEEE
  ICASSP}, vol.~2, 2001, pp. 749--752 vol.2.

\end{thebibliography}

\end{document}